# Time-of-flight recoil detection in transmission using pulsed keV ion beams enables sensitive multi-element profiling with high depth resolution

R. Holeňák[1*], S. Lohmann[1,2] and D. Primetzhofer[1]

[1]*Department of Physics and Astronomy, Uppsala University, Box 516, S-751 20 Uppsala, Sweden*
[2]*Institute of Ion Beam Physics and Materials Research, Helmholtz-Zentrum Dresden-Rossendorf e.V. (HZDR), 01328 Dresden, Germany*

**Abstract**

The potential of time-of-flight recoil detection in transmission geometry using pulsed keV ion beams for sensitive multi-element profiling of thin membranes and quasi-2D systems is assessed. While the time-of-flight approach allows for simultaneous detection of multiple elements, to the largest extent irrespective of recoil charge states, the keV projectile energies guarantee high recoil-cross sections yielding high sensitivity at low dose. We demonstrate the capabilities of the approach using $^{22}$Ne and $^{40}$Ar as projectiles transmitted through thin carbon foil featuring optional LiF-coatings and single-crystalline silicon membranes for different sample preparation routines and crystal orientations.
For a large position sensitive detector (0.13 sr), a high-depth resolution below 6 nm and sensitivity below $10^{14}$ atoms/cm$^2$ was achieved. For crystalline targets, we show how the probability of creation and detection of recoils and their observed angular distribution depend on sample orientation.

*Corresponding author: radek.holenak@physics.uu.se

1. **Introduction**

Energetic ion beams enable a set of tools for quantitative materials analysis, which comes with many benefits such as limited need for sample preparation, simple and well-understood interaction principles enabling highly accurate data evaluation and, last but not least, virtually non-destructive analysis [1]. Keeping these advantages, forward scattering of target recoils can detect light constituents down to hydrogen [2]. Developments such as detector telescopes for simultaneous measurements of specific energy loss and particle energy [3] or time-of-flight and energy [4] allowed to establish elastic recoil detection analysis (ERDA) as a methodology capable of simultaneous detection of all target constituents in a single measurement and a single detector system in particularly using heavy ions to increase cross-sections and achieve more favourable kinematics [5–10].

While initially performed at energies of beyond 100 MeV [11], the energies commonly employed in heavy ion elastic recoil detection analysis (HIERDA) are nowadays often in the range of only a few ten MeV, as for all ion beam methods low energy yields increased sensitivity and allow for using smaller accelerators [12,13]. These lower energies, however, often result in limited probing depth and depth resolution relative to the probing depth [14,15]. With the commonly employed detection geometry of close to 45 degrees with respect to the primary beam, the depth resolution is moreover often challenged by geometrical straggling [15] due to the prolongation of incoming and exiting trajectories of projectile and recoil, respectively.

In contrast, the recoil path length in transmission geometry decreases with the probed depth. The final energy spread of the recoils created on the entry surface is a result of the individual straggling of the recoil in the sample, whereas the recoils formed at the exit surface will all have an energy spread defined by the straggling of the projectile. Several groups have employed the transmission approach using a wide range of beam energies and projectiles masses [16–18]. High sensitivity has been achieved by coincidence ERDA in transmission geometry enabling detection of specific isotopes [19] and has been employed for detection of subtle amounts of hydrogen [20], and also for in-operando profiling of Li in thin film batteries [21–23].

Medium energy ion scattering (MEIS) is known for its high depth resolution and sensitivity achieved by employing projectiles with keV energies. Copel et. al showed that the method can also be used for the detection of light recoils with extreme depth resolution and sensitivity [24]. In what they referred to as MEIS-ERDA singly charged beams of He and Li were used to detect monolayers of hydrogen on epitaxially grown Si crystals [23,24].

In this contribution, we demonstrate time-of-flight elastic recoil detection performed in transmission experiments using a pulsed beam of ions with keV energies. We employ heavier projectiles than previously used in MEIS-ERDA namely Ne and Ar, that allow for simultaneous detection of all light constituents. Moreover, a large position-sensitive detector makes it possible to study the angular distribution of detected recoils. The specific target systems, i.e. amorphous carbon foils and single-crystalline silicon membranes, were chosen for the demonstration of the method and some of its possible applications. Amorphous carbon foils are a vastly used material in accelerators and particle time-of-flight detectors as a stripping medium [27,28]. Hence, they are available in a suitable range of thicknesses while presenting common surface contaminations, which can be assessed in the present approach. Single-crystalline targets allow studying how the periodicity of a crystalline structure affects the trajectories of the projectiles [29] and, consequently, may alter the conditions for recoil creation. By performing experiments in transmission through self-supporting thin films, we can reduce the detrimental effects from geometrical straggling while achieving a high depth resolution and sensitivity due to the drastically increased cross-sections in comparison to common HIERDA approaches. We present the methodology of the transmission approach together with an assessment of sensitivity, accuracy in quantification and effects of the target structure.

## 2. Experiment and data evaluation

Experiments were performed employing the time-of-flight medium energy ion scattering (ToF-MEIS) setup at Uppsala University [30,31]. A Danfysik implanter platform can provide a wide variety of atomic and molecular ion beams with energies ranging from 5 to 330 keV for singly charged ions. A 7 m long beamline equipped with several sets of horizontal and vertical slits delivers the projectiles onto the sample in a beam spot restricted to well below (1 x 1) mm$^2$ and a beam divergence better than 0.056°. By employing electrostatic chopping combined with a gating pulse, the incident beam is shaped into pulses with a duration of typically 1-3 ns. The final steady-state current impinging onto the sample is tunable in the range of fA to pA enabling single ion impact within one pulse on average, which allows for transmission experiments with unique discrimination of flight time of ion and secondary particles [32–34].

Samples prepared as thin self-supporting foils are attached to the 6-axis goniometer located in the centre of the scattering chamber. A large, position-sensitive microchannel plate detector (MCP) with a diameter of 120 mm from RoentDek [35] acts as a stop detector. The detector can be positioned in transmission geometry 290 mm from the foils covering a solid angle of 0.13 sr. Collision products are identified by measuring their flight time to the detector. The angular distribution is determined by two perpendicular delay lines [36].

Two carbon foils with nominal thicknesses 8 µg/cm$^2$ and 10 µg/cm$^2$ were stretched over the openings in the metal sample holder from foils floating in a water bath. Before transferring the foils onto the water, while still being attached to the glass substrate, the thicker foil had a thin layer of LiF evaporated onto the surface facing the atmosphere. Thermal evaporation of the LiF powder was performed in a simple high-vacuum coater with the aim of growing the layer as thin as possible.

Single-crystalline Si(100) membranes with a nominal thickness of 50 nm were purchased from Norcada Inc. [37]. Ion beam analysis using MeV primary particles, conducted in our previous work [38], resulted in a thickness of 53 nm and indicated the presence of light contaminants (H, C, O) on both surfaces. One of in total two investigated silicon membrane samples was, prior to measurements, dipped into hydrofluoric acid (HF) solution for 60 seconds in order to remove the native oxide from silicon. The label "Untreated" was assigned to the sample without any equivalent treatment where the surface is expected to have a few nanometres thick native passivation layer of silicon dioxide.

For the systems described above, the projectiles employed as probes were $^{22}$Ne and $^{40}$Ar, respectively. An example of a time-of-flight spectrum recorded employing a primary beam of 143 keV $^{22}$Ne$^+$ on the 10 μg/cm$^2$ carbon foil with the LiF layer is depicted in figure 1. The fastest products of the ion-solid interactions are photons, arriving at the detector 1 ns after the collision [32]. With a characteristic delay, target recoils arrive starting with hydrogen, followed by lithium, carbon, oxygen and fluoride. Positions of the last two recoil peaks are suggested by arrows, as the O and F recoil peaks are obscured due to the overlap between the low energy tail of the carbon recoil peak and the dominant peak of transmitted Ne projectiles. At longer flight times, the spectra show a higher background formed by the projectiles experiencing high large angles and multiple scattering.

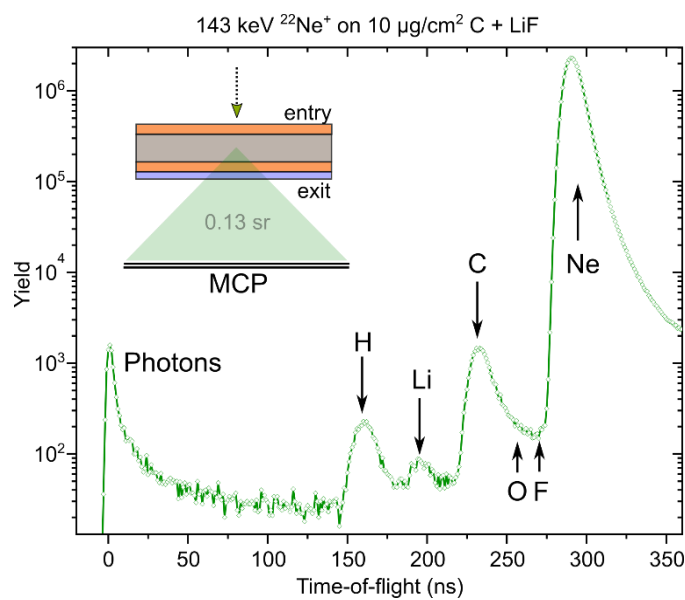

Figure 1: Time-of-flight spectrum recorded for 143 keV $^{22}$Ne$^+$ directed on a 10 μg/cm$^2$ carbon foil with a thin LiF layer evaporated on the exit side. The scheme in the inset shows the layered profile of the sample.

Any real sample investigated in such transmission experiments can be considered to feature a sandwich-like profile with a matrix of finite thickness and two surface layers to which surface adsorbents or reactants such as hydrocarbons or oxygen are expected to contribute (see inset of figure 1). To predict the position of a certain recoil species in the ToF spectra as well as to quantify their concentration, two factors need to be considered on top of scattering kinematics, namely the energy loss of the projectile and recoiling species in the target and the recoil cross-section.

Specifically, recoils created at the surface at which the projectile enters the target need to travel through the entire sample to reach the detector and are, therefore, subject to energy loss in the bulk material. On the other hand, recoils produced at the exit surface are separated from the detector only by the drift region. The momentum of the projectile is, however, decreasing with distance from the entrance surface since the projectile also experiences an energy loss along its trajectory. The energy and thus the detected time-of-flight of recoils originating from different depths is therefore highly dependent on the respective projectile and recoil species and their respective stopping powers in the given material.

Additionally, in crystalline materials, the stopping power was shown to be trajectory dependent [38,39]. In case of the silicon sample, if the sample is positioned in such a way that the [100] crystal axis is aligned parallel to the incident beam, a large majority of projectiles will be channelled. Ions experience only small scattering angles around the incident beam position and are subject to both reduced energy loss and reduced energy loss straggling in comparison to randomly oriented crystals. When rotating the sample by $\theta x$ = 6° around the x-axis and by $\theta y$ = 12° around the y axis, the incident beam is no longer aligned with any low-index crystal axis, and we refer to this geometry as (pseudo)-random. Ions scattered into larger angles are further influenced by the crystalline structure on their outward trajectory, i.e. they are subject to blocking and planar channelling.

Within the range of recoiling angles covered by the solid angle of the detector, the recoil cross-section exhibits a very flat rise. The maximum difference between the centre and the edge of the detector was for all projectile–recoil combinations found to be less than 6 %. A single recoil cross-section value at half of the maximum deflection angle was therefore used for the calculation. At employed projectile energies the recoil cross-section deviates from the Rutherford formula and screening needs to be considered. We used the calculator embedded in the SIMNRA software [40], which provides values of recoil cross-sections calculated applying the screened ZBL potential [41]. At 150 keV Ar, the Rutherford recoil cross-section is already 60 % larger than when considering ZBL screening. In the multilayer structure of the sample (two surfaces and bulk), the characteristic recoil cross-section $\frac{d\sigma_R(E,\theta_R)}{d\Omega}$ was defined for individual layers. At the entrance and exit surfaces, the known incident and measured exit energies were used, respectively, to calculate the recoil cross-section. In the sample bulk, the projectile energy becomes a function of the distance travelled in the material. The resulting recoil cross-section will, therefore, be a function of the energy loss. However, in the limit of small relative energy loss ΔE compared to the projectile energy, the stopping power remains almost constant throughout the projectile entire trajectory in the sample and the recoil cross-section was calculated simply by using the average energy ($E_{in}$+$E_{out}$)/2.

Based on what was stated above, areal densities of recoil species in the target were derived using the following formula based on the single scattering model:

$$N_s = \frac{A}{Q.\Omega.\frac{d\sigma_R(E,\theta_R)}{d\Omega}}$$

Where the quantity A is the integrated area under the recoil peak in the ToF spectra. The integrated charge Q was derived from the absolute number of projectiles arriving at the detector. Such a choice is justified for the cases where the majority of the incident ions undergo only small-angle scattering and are thus covered by the solid angle Ω of the detector. In the range of studied species and energies, the response of the MCP detector is expected to be uniform and since detection efficiency (limited by the open area ratio) is expected to be identical for both recoils and projectiles, it cancels out in the calculations of areal densities. Due to the ToF approach, the charge states of the transmitted projectiles and recoils strongly deviating from unity in the present energy regime [42] are irrelevant, which is beneficial for accurate quantification. In all studied systems, the projectiles directed towards the detector accounted for at least 90 % of all primary projectiles and the lost counts were reconstructed by fitting the radial distribution profile.

All expected target recoils are leaving the thin membrane with higher velocity than the incident ions. As they arrive at the detector before the projectiles, the yield of the identifiable recoil peaks can be straightforwardly integrated. A single scattering model was used for the calculation of the areal densities i.e. every recoil was assumed to be a product of a single collision between a target atom and the incident projectile. Even though recoil cascades do occur and give rise to recoil creation, their contribution to the final recoil peak area is small in comparison to the recoils created by single collisions. Moreover, some of these recoils created as a product of multiple collisions would inevitably suffer higher energy loss and would arrive at the detector with time delay, and thus contribute only to the background. By the same reasoning, the number of projectiles lost in the creation of, in particular heavy recoils, was ignored in the calculation. The range of detectable recoils can, furthermore, be tuned by the choice of the projectile and initial energy. When quantifying recoil yields, the slightly rising background towards higher flight times was fitted by an exponential curve. The steady count rate on the MCP was kept around 10 000 Hz and the spectrum for all recoils was acquired in about 30 minutes ensuring good statistics. The absolute number of projectiles i.e. the dose did not exceed $10^8$ during the entire measurement.

### 3. Results

By examining the ToF spectra in figure 1, one can reason the following: The peak associated with hydrogen recoils is a result of an overlap of two hydrogen recoil distributions from the entrance and the exit surface. In order to calculate the hydrogen areal density, we assume the two surfaces to be equivalent. The lithium recoils originate from a single surface onto which LiF was deposited. Carbon recoils are produced along the entire trajectory of the projectiles and thus peaking above all other detected recoils. Reduced mass separation together with growing overlap between carbon peak and raising projectile peak, obscures detection of oxygen and fluorine, which are expected to accompany hydrogen contaminations and LiF deposition. Nevertheless, since there is no sign of any shoulder on the right slope of the carbon peak, where the oxygen peak is expected, an upper limit for the areal density of oxygen can be estimated.

A reconstruction of the sample composition from the recoil yields for the carbon foils is shown in table 1. The resulting thickness of the carbon foils is in rather good agreement with nominal values. Both foils show high hydrogen contamination. Lithium recoils confirm the successful deposition of the LiF layer. A thickness of about 4.2 nm can be estimated by assuming a density of 2.64 g/cm³ for LiF. The oxygen areal density is estimated to be not higher than $15*10^{15}$ atoms/cm². The final uncertainties are calculated as a combination of statistical error and error introduced in defining the integration window. Sensitivity better than $5*10^{14}$ atoms/cm² and $5*10^{15}$ atoms/cm² could be estimated for H and Li, respectively. The latter would be further improved in the absence of H.

*Table 1: Areal densities and corresponding thicknesses estimated for the two carbon foils with nominal thicknesses of 8 µg/cm2 and 10 µg/cm2. The thicker foil had thin LiF film evaporated onto the exit surface*

| Recoil | 150 keV $^{22}$Ne$^+$ | 143 keV $^{22}$Ne$^+$ |
|---|---|---|
| | $10^{15}$ atoms/cm² | |
| | 8 µm/cm² C | 10 µm/cm² C + LiF |
| C | 481.2 ± 24.6 | 681.6 ± 34.4 |
| C (nm) | 42.3 ± 2.2 | 60.0 ± 3.0 |
| H | 23.4 ± 1.3 | 30.5 ± 1.6 |
| Li | x | 22.5 ± 3.5 |

In the experiments with single-crystalline silicon membranes, we distinguish between two orientations of the beam: channelling and random. Both recoil spectra for the HF-dipped sample using 220 keV Ar are shown in figure 2. Light recoils of H, C and O together with target Si recoils can be identified in both spectra. Nevertheless, a number of differences in peak positions and respective yields between the two spectra are observed. Projectiles detected in channelling geometry have shorter flight times than those detected in random. Moreover, their flight time distribution, and therefore the energy distribution, is much narrower. The flight time shift in the position of the projectile peaks in the spectra is due to a combination of different energy losses and due to sample–detector distance, which increases for the random orientation as a result of changing the crystal rotation. As a direct consequence of these effects, a similar flight time shift is observed for all recoils, but most notably for hydrogen. Recoiled H atoms originating from opposite surfaces are detected as two individual peaks. In random orientation, the variation in stopping and kinematics leads to a repositioning of the peaks and smaller separation. Nevertheless, the two-peak structure can still be distinguished. In the flight time range of 185-255 ns recoils of carbon and oxygen are expected to be found. The small peak at about 200 ns in channelling geometry was identified as a pile-up of two carbon peaks from each of the surfaces arriving at the detector at almost the same time. The same effect is true also for the oxygen recoils, but the detected signal does not extend above the background. In random orientation, the carbon and oxygen recoil peaks do not pile up, instead, the flight time of slower carbon recoils coincides with the faster oxygen recoils obscuring the identification of individual contributions. The most pronounced difference between channelling and random orientation is found for the silicon recoil yield. In random orientation, large amounts of silicon recoils are created along the projectile trajectory directly proportional to the membrane areal density. In channelling, only a fraction of silicon recoils expected to be mainly originating from the outermost layers is detected and again showing a two-peak structure.

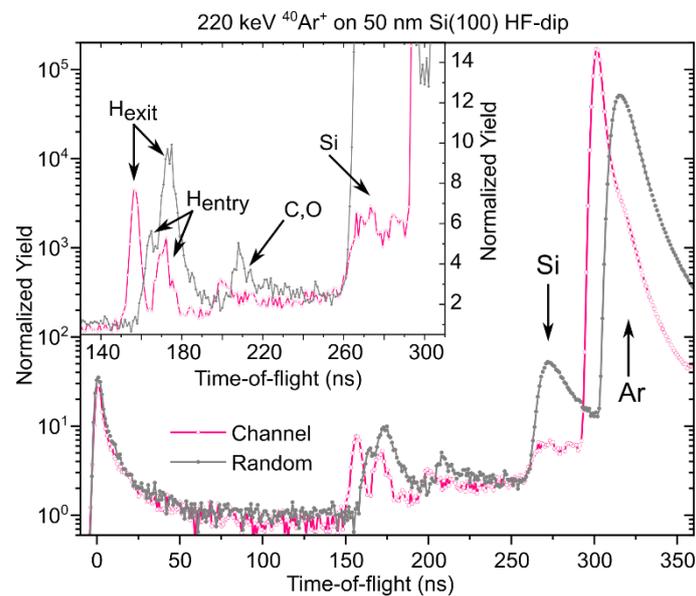

*Figure 2: Comparison of channelling and random spectra acquired from the HF-dipped silicon membrane with a nominal thickness of 50 nm. $^{40}Ar^+$ with an initial energy of 220 keV was employed as a probe. The inset of the figure shows a ToF spectrum focused on all detected recoils with a linear scale.*

Figure 3 shows recoil spectra recorded for the HF-dipped sample from figure 2 in comparison to one that was not subject to equivalent surface treatments. The data was recorded using 150 keV Ar. The initial energy reduced by 70 keV effectively increases the recoil cross-section while still conserving the peak separation. Four recoil peaks of light contaminants are clearly distinguishable in figure 3 - two for individual layers of hydrogen and one each for the piled-up carbon and oxygen, originating from the HF-dipped and the untreated sample, respectively. Here, the slight offset between projectile flight time is attributed to a small difference in sample thickness. From the recoil yields, the effect of the surface treatment can be directly examined. The suppression of the oxygen peak and the decrease in silicon signal points to the successful removal of the native oxide layer, whereas, the signal from hydrogen doubles after the HF-dip. The appearance of an additional carbon signal for the HF-dipped sample suggests the attraction of carbon-based compounds by the surface saturated with hydrogen.

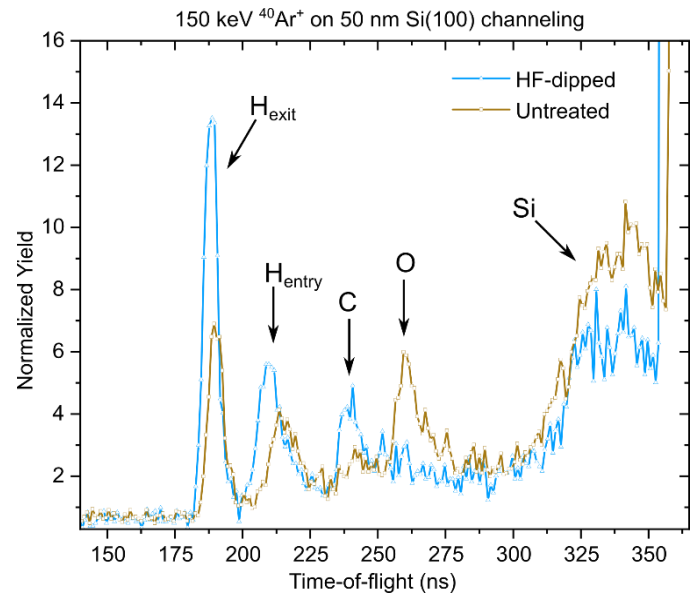

*Figure 3: Channelling spectra of a silicon membrane with and without the HF treatment acquired in the experiment using 150 keV $^{40}$Ar$^+$.*

The results of the quantitative evaluation of the recoil yields, for the silicon membranes with different surface modifications, measured with Ar at two impact energies and two beam-crystal alignments are summarized in table 2. In general, good agreement is found for the two projectile energies employed. Areal densities of light contaminants found on each of the surfaces are in the order of $10^{15}$ atoms/cm². Except for hydrogen in channelling orientation, where each surface contamination could be evaluated individually, the values in table 2 are presented as a contribution from only one of the two equivalent surfaces. The agreement between areal densities derived from the two individual hydrogen recoil peaks found throughout the table supports the validity of the assumed sample symmetry. A thin disordered layer of silicon dioxide can be found at the entrance and exit of the untreated membrane. On the HF-dipped sample, this layer is reduced which comes along with a significant decrease of oxygen signal. In random orientation, the thickness of the silicon membrane in nm is calculated using the silicon density of 2.32 g/cm³. While the thickness of the untreated sample is larger than expected, an excellent agreement is found with the nominal and independently measured thickness of 50 nm and 53 nm, respectively for the HF-dipped sample. A notable discrepancy is found between the amount of hydrogen measured in channelling and random geometry, where the latter shows a consistently higher amount of hydrogen.

*Table 2: Composition of silicon membrane measured with $^{40}Ar^+$ at two incident energies, for two samples with different surface treatment and in two different crystal orientations. Unless noted otherwise in the subscript of the recoil name, the quantities are referring to only one surface layer of the sample. The final uncertainties were calculated as a combination of statistical error and error introduced in defining the integration area. The latter could be in certain cases as high as 25 % due to the simplified background subtraction and peak overlap.*

*\* The summed areal density is thus provided for carbon and oxygen where the contribution of individual peaks could not be separated.*

|  |  | 150 keV $^{40}Ar^+$ | | 220 keV $^{40}Ar^+$ | |
|---|---|---|---|---|---|
|  |  | $10^{15}$ atoms/cm$^2$ | | $10^{15}$ atoms/cm$^2$ | |
| Orientation | Recoil | Untreated | HF-dipped | Untreated | HF-dipped |
| Channel | Si | 13.8 ± 1.4 | 8.9 ± 1.4 | 16.2 ± 2.5 | 8.0 ± 1.6 |
|  | $H_{exit}$ | 4.5 ± 0.3 | 8.9 ± 0.5 | 4.4 ± 0.3 | 8.7 ± 0.5 |
|  | $H_{entry}$ | 4.3 ± 1.1 | 7.6 ± 1.5 | 4.3 ± 1.1 | 7.6 ± 1.5 |
|  | C | 1.4 ± 0.4 | 4.3 ± 0.7 | 1.6 ± 0.4 | 3.6 ± 0.9 |
|  | O | 7.0 ± 1.8 | 1.7 ± 0.4 | 7.6 ± 1.9 | 1.9 ± 0.5 |
| Random | Si | 320.6 ± 32.1 | 268.3 ± 13.5 | 308.2 ± 30.9 | 263.4 ± 13.3 |
|  | Si (nm) | 64.4 ± 6.4 | 53.9 ± 2.7 | 61.7 ± 6.2 | 52.9 ± 2.7 |
|  | H | 6.1 ± 1.3 | 14.2 ± 2.8 | 5.8 ± 0.9 | 13.1 ± 2.0 |
|  | C+O* | 7.5 ± 1.9 | 4.6 ± 1.2 | 7.8 ± 1.6 | 4.8 ± 1.2 |

Complementary to the evaluation of ToF spectra, we have studied the angular distribution of the detected recoils. Figure 4 shows the angular intensity distribution of both transmitted projectiles and silicon recoils extracted from the spectra presented in figure 2. The maps are produced by selecting a flight time range associated with the detection of a specific mass and by reconstructing the respective detection position. The highly ordered structure of the silicon crystal creates an anisotropy in the angular scattering distribution of the projectiles known as a blocking pattern. Lines and circular areas with reduced intensity are projections of crystal planes and channels, respectively, forming a real space image of the silicon crystal structure. The same effect is observed for the silicon recoils created in the target bulk. A clear difference is observed in the flux distribution. While projectile scattering leads to an angular distribution peaking in the initial direction of the beam, the recoils are, apart from the blocking, distributed homogeneously.

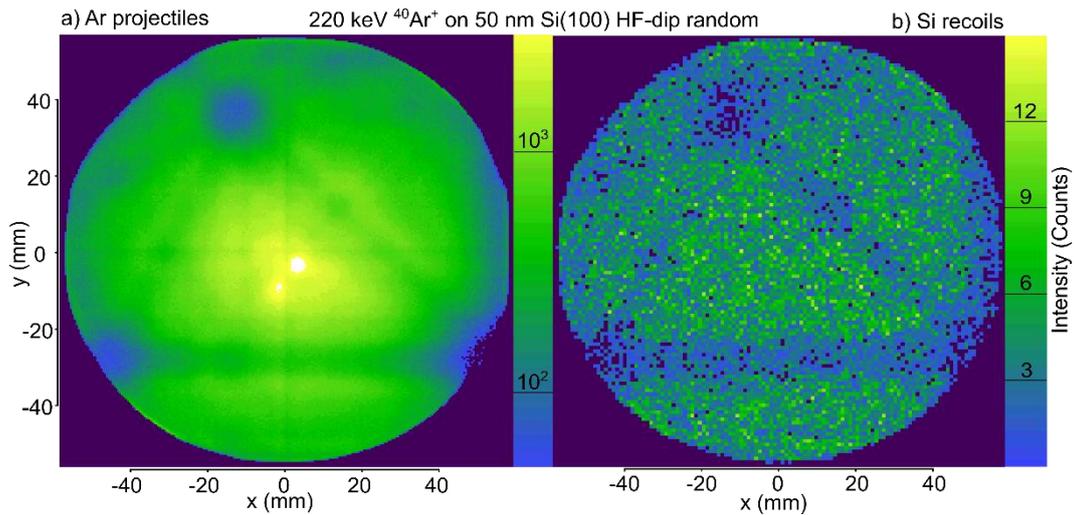

*Figure 4: Angular distributions for simultaneous detection of a) transmitted $^{40}Ar^+$ projectiles with an initial energy of 220 keV and b) recoiled Si atoms from the 50 nm silicon membrane with the beam incident in random orientation.*

In the same manner, the angular distributions of light elements found on the membrane surfaces can be presented. Figure 5 features angular intensity distributions of light recoils along with the transmitted 150 keV Ar projectiles. These distributions are constructed from spectra presented in figure 3 taken in channelling geometry on the HF-dipped sample. The majority of projectiles are detected in the small area in the centre of the detector, in the direction of the beam. Projectiles scattered into larger angles are again subject to blocking, observed more dominantly on the outskirts of the detector. In contrast, the surface recoils form homogeneous maps without any observable structure related to the crystal. The same is true for the light recoils created in random geometry (not shown here).

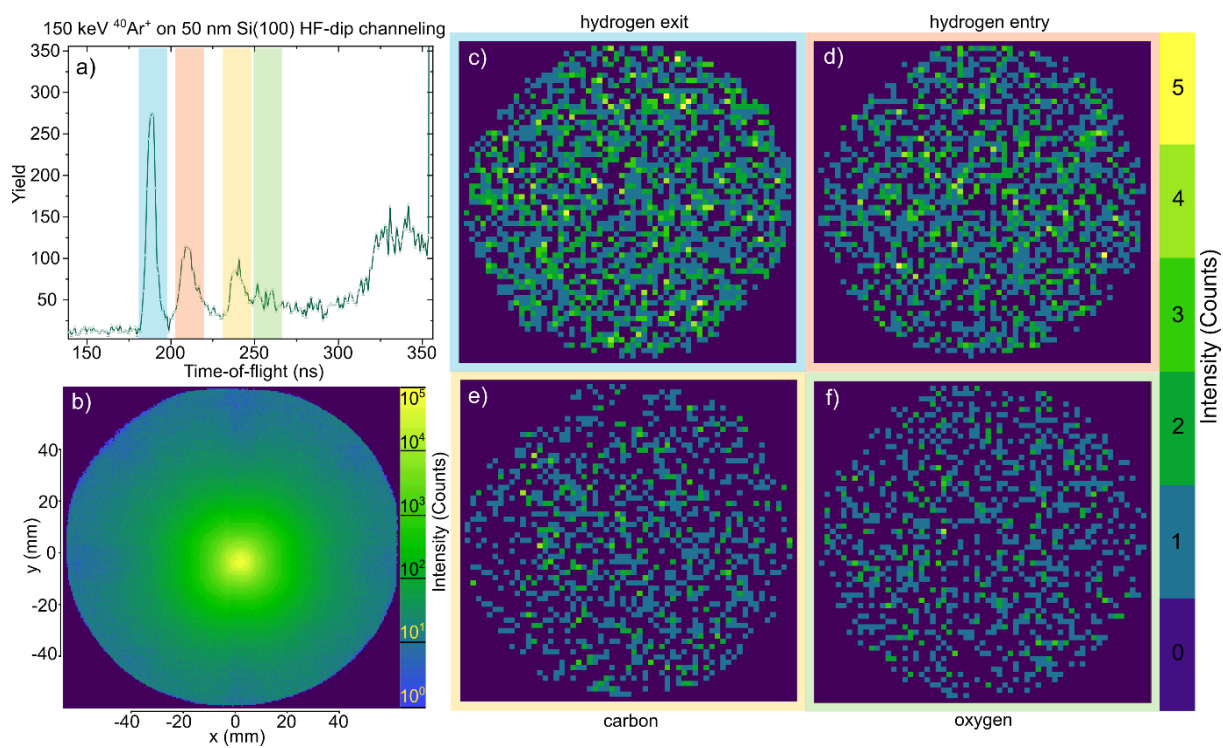

*Figure 5: Angular distributions of b) $^{40}Ar^+$ projectiles with initial energy 150 keV, c) hydrogen from the exit surface, d) hydrogen from the entry surface, e) carbon, f) oxygen with colour-coded frames reconstructed from the regions marked in the a) ToF spectra. Data was acquired from the 50 nm thick HF-dipped silicon membrane in channelling geometry.*

**4. Discussion**

Due to scattering kinematics and energy loss of Ne projectile in the self-supporting carbon foil at given initial energies, the two surface layers of hydrogen become inseparable in the ToF spectrum (see figure 1). Moreover, the energy loss straggling of heavy projectiles in amorphous foils reduces the depth resolution. Nevertheless, the symmetry of the sample and prior qualitative knowledge of the composition allows for the quantification of the areal densities of surface contaminants. For an asymmetrical system, the sample can be positioned so that the grown LiF layer is on the exit surface in relation to the incoming beam. This orientation ensures higher yield due to the larger cross-section and no further straggling of the recoils in the target bulk. Lithium recoil cross-section is the smallest in comparison to all other elements (more than three times lower than for hydrogen). Confirming the presence of a thin LiF layer on the foil is, therefore, a satisfactory result. A considerably high amount of hydrogen on the foil surfaces can be associated with hydrocarbons and adsorbed water – additionally, LiF is hygroscopic.

Transmission of the Ar projectiles through Si in random geometry is analogous to the experiment in the amorphous carbon system. In contrast, channelling could, under ideal conditions, be perceived to be close to two surfaces without a bulk, however, irradiated with projectiles with different energies. Projectiles entering the channel are experiencing only small-angle scattering from atomic strings and are subject to reduced energy loss compared to random geometry. Ideally, no recoils are created along their trajectory inside the crystal channel.

The relative position and separation of the individual recoil peaks from each of the surfaces is a product of the interplay between the, commonly different, specific energy loss of the projectile and the recoil, along with straggling and kinematic factors. The effect of these processes was exemplified when observing the hydrogen recoils in channelling compared to random orientation. A distinct separation of recoil peaks was achieved in channelling.

While this is true for hydrogen in the studied system and for the particular choice of projectiles and energy, it is not a universal rule. Carbon and oxygen recoils from entry and exit surfaces behave differently by reaching the same final energy in channelling geometry but being separated in random orientation. As the recoil cross-section increases for slower projectiles, in a symmetrical system, a larger number of recoils is always expected from the exit surface.

After entering the channel, the projectiles are guided along the crystal axes. Close collisions, resulting in recoil formation, are thus expected to occur mainly on defects or in near-surface layers where channelling has not yet developed and where surface oxidation disrupts the long-distance order. As a result, silicon recoils in channelling geometry are detected as a distribution with a recognizable contribution of two overlapping peaks (see figure 3). On the untreated sample, a fraction of the detected amount of surface silicon can be attributed to the silicon oxide layer. Nevertheless, while the oxygen signal is clearly suppressed in the HF-dipped sample, a substantial amount of silicon remains present in the near-surface layer. Along with the large uptake of hydrogen signal, far beyond what would be considered as a surface monolayer as in the work of Copel et. al [24] on epitaxially grown silicon crystals, we assume that the new surface consists of amorphized silicon saturated with hydrogen and hydrocarbons. This result can be rationalized since no surface cleaning was performed prior to the HF-dip and the original surface structure of the thin membrane is rather unknown. What is intriguing is the disproportionality between the amount of hydrogen detected in different experimental geometries. This observation suggests an ordered structure of the hydrogen bonds, whose detection would be under a certain angle affected by the crystalline structure of the sample. Further experimental work on well-prepared hydrogenated surfaces may help to explain the present data.

While the structural contrast emerging from the intensity distribution of transmitted projectiles such as the one in figure 4a) has been previously reported [43], the blocking pattern observed for forward recoils introduces yet another contrast in ion transmission experiments exploitable e.g. in the fields of transmission ion microscopy [44,45] or in studying the accumulation of implantation damage [46,47]. To our knowledge, this is the first time that recoil blocking is presented in the transmission geometry in the keV energy regime. The recoil blocking effect has been previously used by nuclear physicists to measure nuclear lifetimes [48] and blocking patterns have also been observed using heavy ions with MeV energies. [18]. Our method, however, achieves far superior depth resolution and angular resolution. Also, virtually no damage is being introduced to the sample as the absolute dose accumulated over a single measurement does not exceed $10^8$ of singly charged ions.

No blocking patterns were observed in the angular distributions of the recoils originating from the surface layers in either channelling or random geometry. This observation can be understood as a consequence of these atoms not being aligned with the crystal matrix, i.e. the displacement of the surface contaminants with respect to the crystal channels and rows eliminates the shadowing effect. Additionally, the contribution from the surface facing the detector is obviously not expected to show any blocking.

In the ToF measurement, where the time resolution is defined by the beam pulse width, the final energy resolution of the particles resulting from the collision is mass-dependent. Conversion of the ToF spectra to energy is non-linear and must be done individually for each particle.

For the 150 keV Ar projectiles, the energy resolution resulting from the 2.5 nanosecond pulses is 3 keV, the corresponding resolution for slow hydrogen recoils is about 300 eV. With this resolution, it is possible to resolve the kinematic spreading caused by angular dependence of the energy transfer between the projectile and recoils. Figure 6a) shows a hydrogen recoil peak from the exit surface split into the contribution of fast and slow (more and less energetic) recoils. The corresponding angular distributions are plotted in the maps c) and d). The different angular distributions are apparent as the less energetic recoils are arriving only to the outskirts of the detector. The energy difference between the product $K(0°) \cdot E_{out} - K(11°) \cdot E_{out}$ is for the extreme cases of 0° and 11° degree recoil angle about 500 eV, K being the kinematic factor for hydrogen recoils and $E_{out}$ the measured exit energy of the projectile. It should be noted that the angular maps are already corrected for the flight path difference due to the larger recoil angle. The prolongation of the path length in the sample is negligible. The observed effect is therefore a clear signature of a resolved angular dependency of energy transfer. This feature can be in future utilized to discriminated single scattering events between the projectiles and target atoms e.g. in transmission experiments on 2D materials or in studying crystal surface reconstruction. The only limiting factor is in the moment the pulse width of the incident beam.

Planned improvements of the beamline and pulsing equipment are expected to reduce the pulses down to hundreds of picoseconds. The energy resolution for hydrogen recoils would then be in the range of tens of eV.

Under current conditions, the depth resolution for hydrogen profiling can be estimated from the separation of the entrance and exit hydrogen recoil peaks. An energy spectrum of hydrogen recoils is shown in figure 6b). These recoils were acquired from the centre of the detector within a radius of 20 mm corresponding to a scattering angle of max 4°. The energy separation between the peaks fitted by Gaussians was divided by the averaged 2σ of both peaks. Knowing the absolute thickness of the membrane separating the hydrogen-containing layers, a depth resolution of about 6 nm can be estimated.

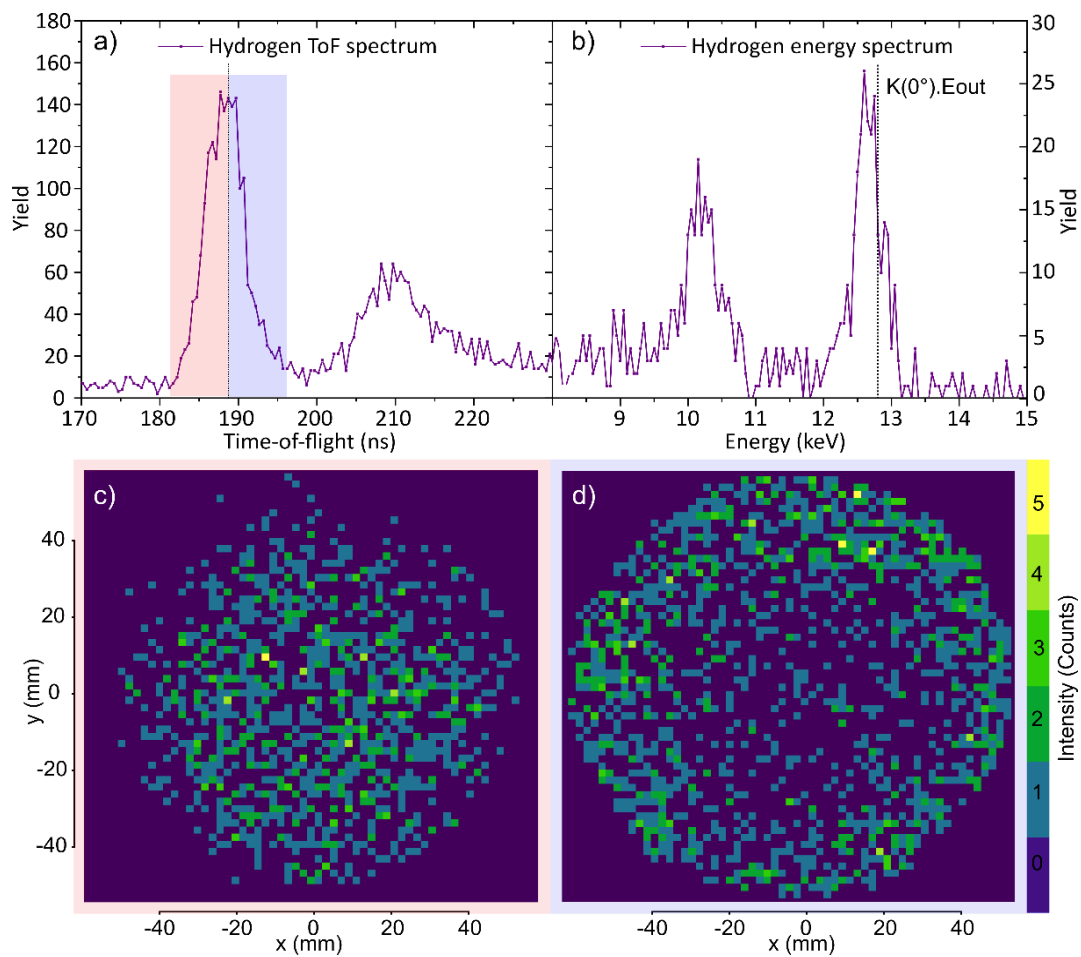

*Figure 6: a) ToF spectra for hydrogen recoils. Angular distributions corresponding to the fast and slow hydrogen recoils produced at the exit surface of the HF-dipped sample by 150 keV $^{40}Ar^+$ are displayed in c) and d) respectively. b) Energy spectra reconstructed from the centre of the detector within a radius of 20 mm.*

Finally, a case can be made for the sensitivity and depth resolution of the present method. In comparison to conventional HIERDA using e.g. 36 MeV $^{127}$I$^{+8}$ at 45°, the recoil cross-section of 7.8 x 10$^7$ mb/sr achieved for hydrogen when using 150 keV $^{40}$Ar$^+$ ions exceeds the former of 5 x 10$^5$ mb/sr by more than two orders of magnitudes. In contrast to the electrostatic analyser (ESA) type detectors [24], the charge states of the recoils are irrelevant although they might deviate significantly from unity [42]. In combination with a large solid angle (0.13 sr) covered by the position-sensitive detector, the method offers a high level of sensitivity and high accuracy. Judging from the peak to background ratio of the exit hydrogen recoil peak portrayed in figure 6 a), the sensitivity is around 10$^{14}$ atoms/cm$^2$. The reduced recoil cross-section for other elements, as compared to hydrogen, leads to a slightly lower sensitivity level, however still in the 10$^{14}$ atoms/cm$^2$ range. This can be truly appreciated when considering the particle beam flux not exceeding 6 × 10$^6$ ions cm$^{-2}$ s$^{-1}$. MeV ERDA with fluxes usually higher by two orders of magnitude is often challenged by undesirable effects induced by the particle fluence on the target like e.g. ion-induced release of light constituents from the sample [49,50], diffusion, amorphisation of crystalline structure and, especially in case of thin foils, irreversible damage by breaking the foil. In contrast, the presented method preserves the composition of the sample and induces virtually no damage. We also want to stress that the accuracy of the quantification is not affected by uncertainties in the stopping power values as these are directly accessible from the measurement itself. The only remaining uncertainties limiting the accuracy of the method are thus found in the scattering potential, the possibility of multiple scattering and, for the moment, the simplified background subtraction.

## 5. Conclusion and outlook

Elastic recoil detection analysis in a transmission time-of-flight approach was demonstrated by characterizing thin self-supporting amorphous and crystalline targets. Deposition of an LiF ultrathin layer on carbon foil was confirmed by detection and quantification of Li recoils produced by the impact of Ne projectiles. The thickness of the layer was quantified to 4.2 nm. Simultaneous detection of other light recoils like hydrogen and carbon allowed for composition analysis of the surface contamination of the sample and estimation of the foil thickness, respectively.

Single-crystalline silicon membranes were studied using Ar projectiles allowing simultaneous detection of a wide range of recoils. Qualitative analysis of the recorded spectra along with the calculated areal densities confirmed an HF-dip induced removal of the silicon oxide layer and consequent saturation of the surface with hydrogen. Differences in the amount of recoils detected in the employed geometries were ascribed to the effect of the crystalline structure on the recoil formation and transport. Moreover, an intensity contrast was observed when studying the angular distribution of the silicon recoils, showing a real space image of the crystal structure. Under the condition employing $^{40}$Ar$^+$ as a projectile, sensitivity in the range of $10^{14}$ atoms/cm$^2$ was achieved for all light elements. The resolution for depth profiling of hydrogen was found to be around 6 nm. Both the sensitivity and resolution can be further improved by tuning the experimental condition for the specific choice of the element of interest.

The versatility in changing the probe and probing energy along with a precise goniometer opens for a wide range of future applications. To mention a few, one can think of in-operando laser annealing, lattice site location exploiting channelling and associated flux enhancement, mapping of the crystal surface reconstruction and characterization of adsorption site of light elements similar to [51], depth profiling of light constituents or studying surface catalytic processes like e.g. hydrosilylation [52]. The capability of studying Li with high depth resolution would permit the study of material diffusion in thin-film stacks to which a voltage is applied, which has been demonstrated in-operando using MeV primary ions [21,22].

An *in-situ* preparation instrumentation connected to the analytical chamber is currently under commissioning. This upgrade to the existing system will allow for complex synthesis and modification under a wide range of conditions. The presented method thus finds its place as a valuable piece in the analytical toolbox for near-surface materials analysis of sensitive self-supporting foils and quasi 2D systems.

**Acknowledgement**

We express our gratitude to Tuan T. Tran for performing the HF-dips.

Accelerator operation was supported by the Swedish Research Council VR-RFI (Contracts No. 2017-00646_9 and 2019-00191) and the Swedish Foundation for Strategic Research (Contract No. RIF14-0053).